\documentclass[12pt]{article}
\usepackage{graphicx}

\def\Title#1{\begin{center} {\Large #1 } \end{center}}
\def\Author#1{\begin{center}{ \sc #1} \end{center}}
\def\Address#1{\begin{center}{ \it #1} \end{center}}

\newenvironment{Abstract}{\begin{quotation}  }{\end{quotation}}
\newenvironment{Presented}{\begin{quotation} \begin{center} 
             PRESENTED AT\end{center}\bigskip 
      \begin{center}\begin{large}}{\end{large}\end{center} \end{quotation}}
\def\Acknowledgements{\bigskip  \bigskip \begin{center} \begin{large}
             \bf ACKNOWLEDGEMENTS \end{large}\end{center}}
             
\newcommand{\adeg}{\ensuremath{^\circ}}             

\def\beq{\begin{equation}}
\def\eeq#1{\label{#1}\end{equation}}
\def\eeqn{\end{equation}}

\def\beqa{\begin{eqnarray}}
\def\eeqa#1{\label{#1}\end{eqnarray}}
\def\eeqan{\end{eqnarray}}

\let\bar=\overbar

\def\etal{{\it et al.}}

\def\Dslash{\not{\hbox{\kern-4pt $D$}}}
\def\dslash{\not{\hbox{\kern-2pt $\del$}}}

\def\msb{{\bar{\ssstyle M \kern -1pt S}}}

\begin{document}
\begin{titlepage}

\vfill
\Title{Surveying the TeV sky with HAWC}
\vfill
\Author{Robert J. Lauer}
\Address{Department of Physics and Astronomy, University of New Mexico, 
Albuquerque, NM 87131, United States of America\\
rlauer@phys.unm.edu}
\Author{for the HAWC Collaboration}
\Address{
For a complete author list, see 
http://www.hawc-observatory.org/collaboration/icrc2015.php}
\vfill
\begin{Abstract}
The High altitude Water Cherenkov (HAWC) Observatory has been completed and 
began full operation in early 2015. Located at an elevation of 4,100 m near the 
Sierra Negra volcano in the state of Puebla, Mexico, HAWC consists of 300 water 
tanks instrumented with 4 PMTs each. The array is optimized for detecting air 
showers produced by gamma rays with energies between 100~GeV and 100~TeV and 
can also be used to measure charged cosmic rays. A wide instantaneous field of 
view of $\sim 2$ steradians and a duty cycle $>95$\% allow HAWC to survey 
two-thirds
of the sky every day. These unique capabilities make it possible to monitor 
variable gamma-ray fluxes and search for gamma-ray bursts and other transient 
events, providing new insights into particle acceleration in galactic and 
extra-galactic sources. In this contribution, we will present first results from 
more than one year of observations with a partial array 
configuration. We will discuss how HAWC can map the gamma-ray sky as well as 
probe other physics including cosmic ray anisotropies and the search for 
signatures of dark matter annihilation. 
\end{Abstract}
\vfill
\begin{Presented}
Twelfth Conference on the Intersections of Particle and Nuclear Physics 
(CIPANP)\\
Vail, Colorado, United States of America, May 19--24, 2015
\end{Presented}
\vfill
\end{titlepage}
\def\thefootnote{\fnsymbol{footnote}}
\setcounter{footnote}{0}

\section{Introduction}

Gamma-ray astronomy at TeV energies has opened a new window into the 
most energetic processes in our universe, including galactic particle 
acceleration sites like supernova remnants and the extremely variable 
extra-galactic emission from gamma-ray bursts (GRB) and active galactic nuclei 
(AGN). On the one hand, this allows us to probe astrophysical phenomena in 
a new way, and, on the other hand, provides the means to study fundamental 
aspects of TeV-scale physics. 
Bright extra-galactic flares can be used to understand 
cosmological features or test the laws for photon propagation beyond 
energies achieved in laboratories. Furthermore, it is possible to search for 
gamma rays emitted in processes involving hypothetical particles, in particular 
candidates for dark matter, and thus provide invaluable 
experimental constraints for particle physics models. In order to move TeV 
gamma-ray astronomy towards such precision measurements, it is imperative to 
focus more on surveys and statistically analyze classes of objects instead of 
individual case studies. This contribution highlights how the High Altitude 
Water Cherenkov (HAWC) observatory has started to pursue these goals as the 
most sensitive TeV survey instrument.

\section {The HAWC Observatory}

HAWC is an array of 300 water Cherenkov detectors (WCDs). Completed in March 
2015, 300 WCDs housed in commercial steel tanks of $7.3$~m diameter and $4.5$~m 
height cover an area of approximately $22,000~$m$^2$. A light-proof 
bladder in each WCD is filled with $180,000$~liters of purified water. At 
the bottom, three $8''$ photo-multiplier tubes (PMTs) and one central high 
quantum efficiency $10''$ PMT are facing upwards to detect Cherenkov light from 
relativistic particles, produced as secondaries in extensive air showers.
HAWC is located at an altitude of $4,100$~m above sea level and thus closer to 
where the maximum number of secondary particles occur during shower evolution 
than the predecessor MILAGRO~\cite{bib:MILAGRO}. This results in a lowered 
threshold of $\sim100$~GeV for HAWC and a gain in sensitivity to the steeply 
falling power-law spectra of astrophysical gamma rays.

Signals from every PMT in HAWC are transmitted to a central 
counting house via the same cables that provide the high voltage. Custom 
front-end electronics translate the voltage pulses into digital 
time-over-threshold (ToT) records with sub-nanosecond precision for each time 
stamp.  
The relative timing of all PMTs can be calibrated to nanosecond 
precision with an optical calibration system. Short 
laser pulses of 300~ps are sent via a network of optical fibers into each WCD, 
providing measurements of the delay of the electronic PMT responses to photon 
signals. By varying the laser intensity over four orders of magnitude, the 
system is also used to provide a charge calibration for each PMT.
A trigger, fully configurable via software on the local computers, 
reduces the data rate to about 25 kHz.

Reconstructing the characteristics of the primary shower 
particle for a triggered event includes two steps: first, the charge 
distribution of an event is fitted to determine the central axis of the shower 
(shower core), then, a function describing a curved shower front is fitted to 
the arrival times of all PMT with hits in the event. 
The resulting incident angle is then recorded as the direction of the primary 
particle. The angular resolution improves with the size of the shower 
``footprint'' on the detector, and thus, on average, with the primary's energy. 
In the preliminary analysis, the resolution changes from $\sim2\adeg$ to 
$<0.5\adeg$ over HAWC's energy range and is expected to be as low as 
$\sim0.1\adeg$ for the highest energies with a fully matured analysis.

Most air shower events that HAWC records are induced by hadronic, charged 
cosmic rays. Studying the angular distribution has confirmed and expanded the 
observations of a small-scale (O(10\adeg)) anisotropy in the arrival directions 
at the level of $10^{-4}$ in relative intensity, see~\cite{bib:hawc-cr}. 
The discrimination of events induced by gamma rays from this background relies 
mainly on the fact that the hadronically induced showers typically include 
muons, while gamma-ray showers generally do not. By rejecting events with large 
charge depositions outside the core region, as only expected from muons passing 
near a PMT shielded by $\sim4$~m of water, a large fraction of events induced 
by hadronic primaries can be filtered out.

HAWC gamma-ray analyses are typically performed as binned analyses, 
where events are sorted depending on the fraction of PMTs that are hit by an 
air shower. This variable is a proxy of energy. After applying individual 
gamma-hadron separation cuts in each analysis bin, the reconstructed 
directions of the remaining signal and background events are
collected in a pixelated sky map. The expected background rate at any point can 
then be estimated via direct integration of the background outside the 
region of interest and is used to calculate significances of excesses or 
deficits~\cite{bib:atkins}. The resulting sensitivity of HAWC to 
point-like sources of gamma rays is discussed in detail 
in~\cite{bib:hawcsensi}.

Due to the larger path lengths through the atmosphere for air showers arriving 
from directions closer to the horizon, the sensitivity of HAWC to TeV 
gamma rays is strongly suppressed beyond zenith angles of $45\adeg$. This 
defines an instantaneous field of view of about 2 steradians. Since HAWC is 
operating with a duty cycle $>95$~\%, this wide angular acceptance allows 
HAWC to scan about two-thirds of the entire sky every day.

\section{First Results}

\begin{figure}[htb]
\centering
\includegraphics[width=12cm]{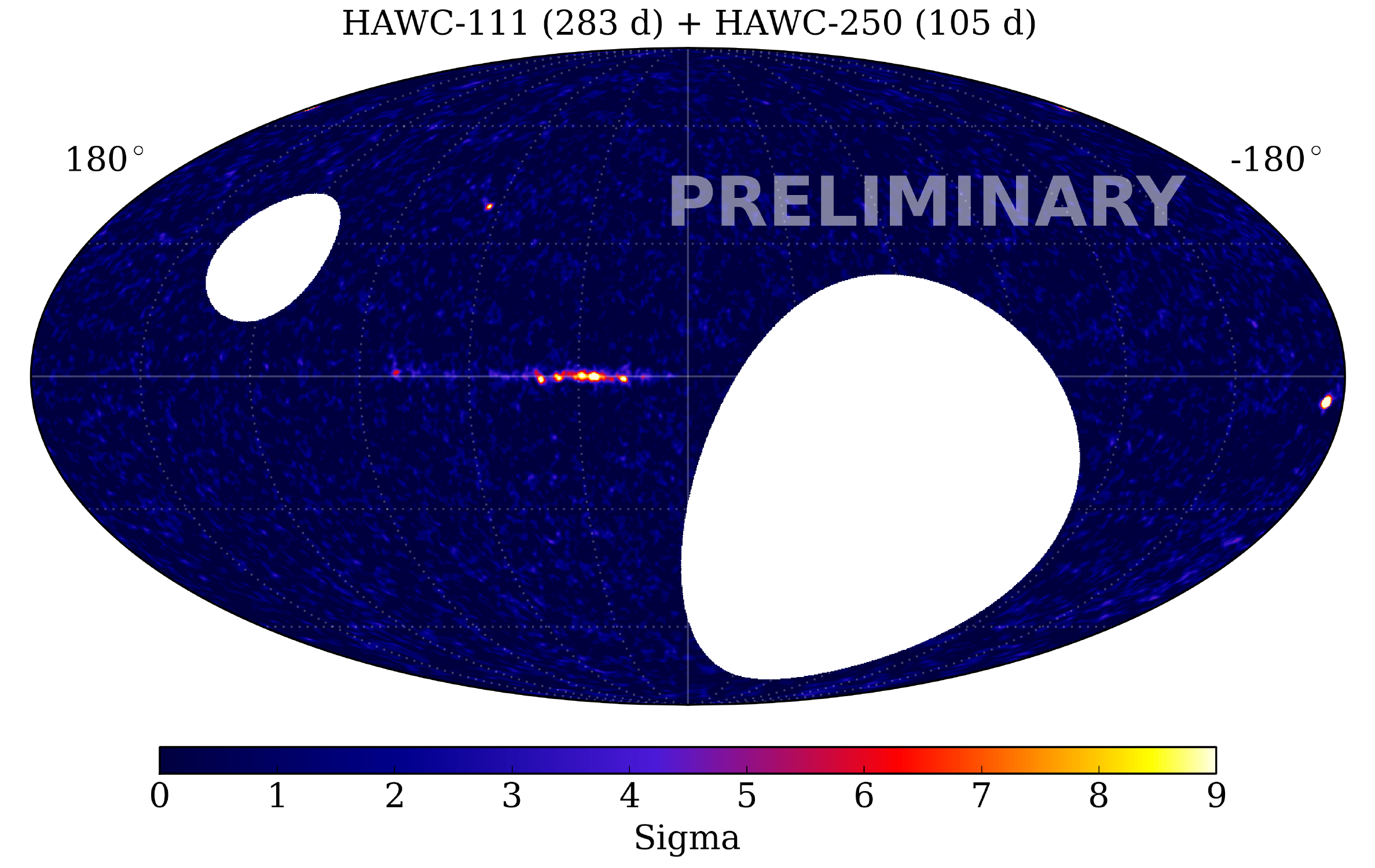}
\label{fig:skymap}
\caption{Point source significance sky map in galactic coordinates for the 
combined data from early operation with about one third of the array 
(\textit{HAWC-111}) and from the period during which the detector grew from 250 
to the final 300 WCDs (\textit{HAWC-250}).}
\end{figure}

During ongoing construction, HAWC took data with an array growing from 90 to 
135 active WCDs between August 2013 and July 2014. In November 2014, a 
configuration of 250 WCDs became operational and the full array of 300 WCDs 
started taking data in March 2015. Figure~\ref{fig:skymap} shows a significance 
sky map of the data from 283 live days of the former configuration (called 
\textit{HAWC-111}) and 105 days of the latter. Among the most prominent 
features is an accumulation of TeV gamma-ray sources along the galactic 
plane. In the inner part of this region, a deconvolution assuming multiple 
point-like sources with overlapping event distributions, due to HAWC's limited 
angular resolution, revealed 10 sources with $>3\sigma$ post-trial significance, 
as discussed in~\cite{bib:galplane}. The brightest spot on the right of the map 
is the Crab Nebula, the brightest steady TeV gamma-ray source. It serves 
partially as a calibration source for HAWC, for example for optimizing the 
gamma-hadron separation cuts~\cite{bib:crab}.

The survey capabilities of HAWC provide an excellent opportunity to perform an 
indirect search for dark matter with HAWC, motivated by models in which 
annihilation or decay of dark matter produce gamma rays. In regions 
where an over density of dark matter is expected, for example around the center 
of our galaxy, a search could reveal such gamma rays if dark matter consists of 
WIMPs 
with masses in the TeV range. A dark matter search in the galactic center is 
complicated by the presence of many baryonic astrophysical sources of 
gamma rays and will require a more detailed mapping of that region first. 
Meanwhile, a simpler TeV dark matter study focusing on dwarf 
spheroidal galaxies that contain no other TeV gamma-ray sources has already 
been performed with early HAWC data. Limits on the dark matter annihilation 
cross section are presented in~\cite{bib:hawcDM}, based on the non-observation 
of significant gamma-ray excesses from 14 dwarf spheroidal galaxies.

\begin{figure}[htb]
\centering
\includegraphics[width=12cm]{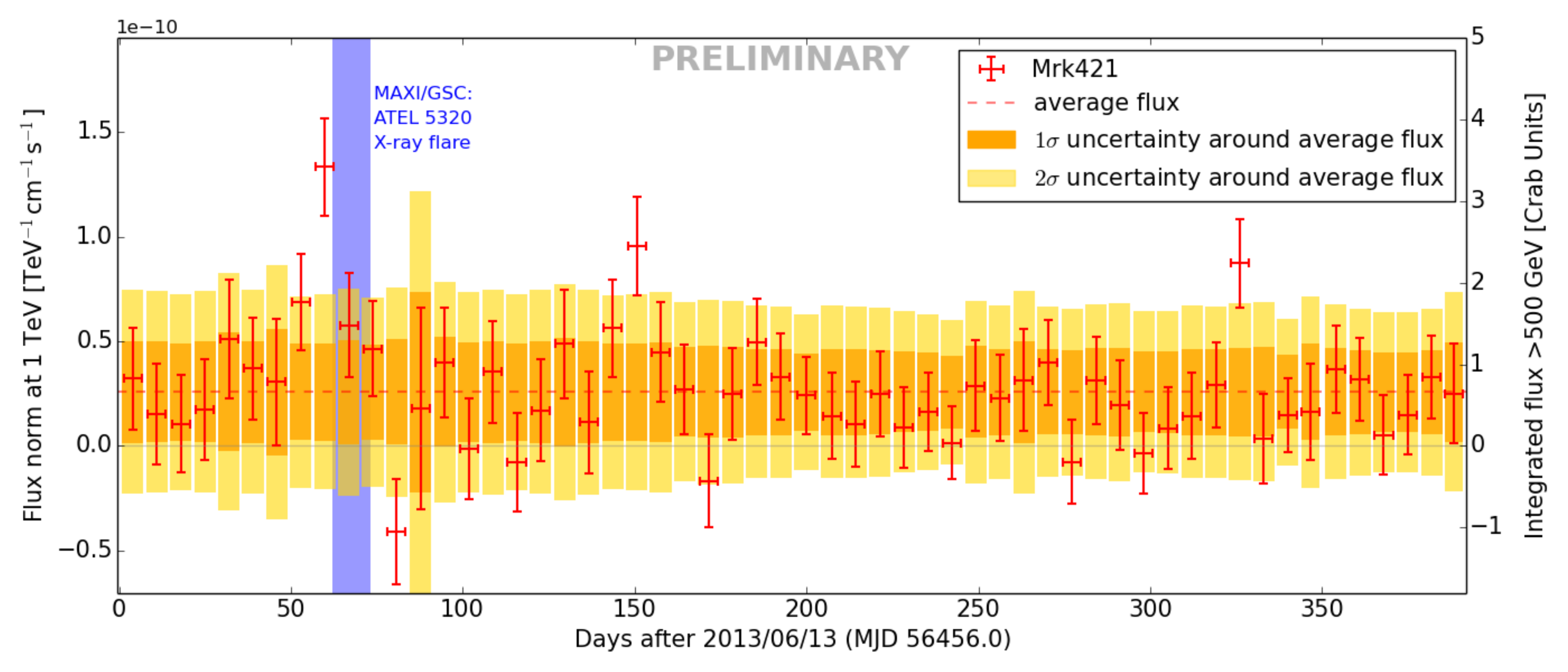}
\label{fig:lc-mrk421}
\caption{Flux light curve of Markarian 421 for the time between June 13, 2013, 
and July 9, 2014, in intervals of 7 days, based on fits with a fixed 
spectral shape with power law index 2.2 and exponential cut-off at 5~TeV and 
are shown as divided by the average Crab flux observed by HAWC on the right 
axis. The large uncertainties in the flux around day $\sim90$ are due to a 
period of construction and maintenance in September 2013 during which HAWC was 
shut down during day time for several days in a row.}
\end{figure}

With its ability to scan two-thirds of the sky every day, HAWC provides 
unparalleled data to search for variability at TeV energies in many different 
objects simultaneously. For the early \textit{HAWC-111} data, light curves of 
bright and highly variable extra-galactic objects were obtained by binning 
the data into 7-day intervals and fitting a gamma-ray flux value for each 
such bin. Figure~\ref{fig:lc-mrk421} shows the light curve for AGN 
Markarian 421. An X-ray flare reported by the MAXI detector happens at the onset 
of the interval in HAWC data with the highest flux 
measurement~\cite{bib:atel5320}. The chance probability of the HAWC measurement 
alone to occur as a fluctuation over a constant average flux, with a trial 
factor for 56 time bins, is $1.4\cdot10^{-5}$ ($4.3\sigma$). This establishes 
HAWC as a tool to monitor day-scale TeV flaring from AGN and these observations 
will be used to constrain acceleration models for such sources. By collecting 
data of extra-galactic gamma-ray emission from many flares and different 
AGN, it will be possible to systematically probe photon propagation and 
attenuation and thus study cosmological features like the extra-galactic 
background light~\cite{bib:EBL} or inter-galactic magnetic 
fields~\cite{bib:igmf}. Further examples and details about this analysis are 
provided in~\cite{bib:blazars} and the same methods will be used to search for 
gamma-ray variability over the whole field of view of HAWC. A separate search 
for much shorter flares (seconds) of TeV gamma rays from GRBs revealed no 
significant excesses in early HAWC data, see~\cite{bib:GRB13,bib:GRBicrc}.

The analyses of the galactic plane sources, dark matter annihilation and light 
curve measurements discussed above were performed within an analysis framework 
developed specifically for the wide range of HAWC science 
topics~\cite{bib:LiFF}. It is based on convoluting model descriptions of 
sources with a detector response, established through simulation and reference 
sources, and comparing those with the observed data via the likelihood 
formalism. A maximum-likelihood procedure can then yield best-fit results for 
the parameters of the models. The HAWC analysis software uses the same 
model interface as that implemented in the more general Multi-Mission Maximum 
Likelihood (3ML) framework~\cite{bib:3ML}. The latter software is designed 
to support direct joint analyses of data from different experiments. This 
feature will allow us to perform studies over a much wider energy range than 
HAWC alone covers, for example by including Fermi-LAT data in a joint spectral 
fit for any of the above source classes.

\section{Conclusions and Outlook}

The HAWC observatory has been continuously taking data for two years, first 
with a partial configuration and then with the completed array since March 
2015. The wide field of view and $>95$~\% duty cycle provide unprecedented 
observations of TeV gamma-ray emission from galactic and extra-galactic 
sources and their spectra and variability. First results include a scan of 
dense clustering of gamma-ray sources in the galactic plane, a search for TeV 
dark matter annihilation and the first day-scale TeV light curves obtained with 
a wide-field water Cherenkov detector. More sensitive analyses based on the 
most recent data taken with the completed HAWC array will soon reveal many new 
insights into the TeV sky.

\Acknowledgements

\section*{Acknowledgments}
{\footnotesize
We acknowledge the support from: the US National Science Foundation (NSF);
the US Department of Energy Office of High-Energy Physics;
the Laboratory Directed Research and Development (LDRD) program of
Los Alamos National Laboratory; Consejo Nacional de Ciencia y Tecnolog\'{\i}a 
(CONACyT),
Mexico (grants 260378, 232656, 55155, 105666, 122331, 132197, 167281, 167733);
Red de F\'{\i}sica de Altas Energ\'{\i}as, Mexico;
DGAPA-UNAM (grants IG100414-3, IN108713,  IN121309, IN115409, IN111315);
VIEP-BUAP (grant 161-EXC-2011);
the University of Wisconsin Alumni Research Foundation;
the Institute of Geophysics, Planetary Physics, and Signatures at Los Alamos 
National Laboratory;
the Luc Binette Foundation UNAM Postdoctoral Fellowship program.
}

\end{document}